\documentclass[twocolumn,showpacs,preprintnumbers,amsmath,amssymb]{revtex4}

\usepackage{graphicx}
\usepackage{dcolumn}
\usepackage{bm}

\begin{document}

\title{Unstable electromagnetic modes in strongly magnetized plasmas}

\author{S. Son}
\affiliation{18 Caleb Lane, Princeton, NJ 08540}
\author{Sung Joon Moon}
\affiliation{28 Benjamin Rush Lane, Princeton, NJ 08540}
\date{\today}

\begin{abstract}
The electromagnetic modes possibly 
unstable in  strongly magnetized plasmas 
are identified. 
The regime where  this instability might stand out
compared to the incoherent electron-cyclotron radiation is explored.  
These modes are relevant to
the inertial confinement fusion and the gamma ray burst.
\end{abstract}


\maketitle

\section{Introduction}
The magnetic field is ubiquitous in plasmas.
The plasma interaction with the magnetic field often determines 
its dynamical behavior~\citep{Stone, Biskamp, Fishman, Nakar, Innes}.
In particular, the magnetic field could convert
the electron kinetic energy into the collective photons,
as in the magnetron or the gyrotron~\citep{gyrotron3, tgyro}
and the free electron laser (FEL)~\citep{Fisch, sonlandau, sonbackward}. 
In the FEL, a relativistic electron beam 
 gets periodically  accelerated amplifying  coherent 
 electromagnetic (E\&M) waves.  
The resulting lasers have various applications.
One natural question would be
whether the electron thermal gyro-motion in the magnetic field, not the specifically designed population-inverted plasma state,  
can excite an analogous process.

In this letter, we examine the instability of collective E\&M waves arising 
from the electron thermal gyro-motion.
The analysis of the electron motion in the presence of a strong magnetic field
leads to a theoretical framework similar to that of the Landau damping~\citep{sonlandau, songamma},
which reveals that there could exist numerous unstable E\&M modes in  relativistic plasmas.
Our theory is similar to the well-known electron maser theory~\citep{chu, Treumann}, but our theory deals with the long time limit
while the previous researches mainly focus on the short time limit, which 
 will be discussed in detail at the end of Sec.~(4).  
We identify the regime where the coherent radiation from this instability 
might be  more intense than the incoherent cyclotron radiation. 
Various implications of our study on the astrophysical and laboratory plasmas are discussed,
including the short gamma ray burst~\citep{Nakar}, the non-inductive current drive~\citep{Fisch, sonprl} and 
the soft x-ray generation in the inertial confinement fusion~\citep{Tabak}.  

This paper is organized as follows. 
In Sec.~(2), the theory of the instability is developed based 
on the Landau damping theory for the non-relativistic plasmas.
In Sec.~(3), 
the fully relativistic theory is developed.   
This section is the major result of this paper.  
In Sec.~(4), we identify the instability inherent in the relativistic plasmas.  In Sec.~(5), we discuss the implications of our results.  

\section{ Landau theory: Non-relativistic electrons}
In this section, we consider the non-relativistic plasma. 
While we will conclude that the theory developed in this section is inadequate,
the intuition derived from this development 
 is  useful for the fully relativistic theory  in the next section.  
We start with a non-relativistic electron
under the magnetic field $\mathbf{B}_0 = B_0 \hat{z}$. 
The equation of motion is
\begin{equation} 
m_e\frac{d \mathbf{v}}{dt}= -e \left( \mathbf{E}_0 + \frac{\mathbf{v}}{c} \times  \mathbf{B}_0 \right) \mathrm{,} \nonumber 
\label{eq:motion}
\end{equation}
where  $m_e$ ($e$) is the electron mass (charge), $ \mathbf{E}_0= 0 $,  $\mathbf{v}$ is the electron velocity,
and $c$ is the speed of light. 
The zeroth-order solution is $v^{(0)}_{x}(t) = v_p \cos(\omega_{ce} t +\phi_0)$,
$v^{(0)}_{y}(t) = -v_p \sin(\omega_{ce} t +\phi_0)$ and $v^{(0)}_{z}(t) = v_{0z}$, where $\omega_{ce} = eB_0/m_e c $,
$v_p$ is the constant perpendicular velocity,
and $\phi_0$ is the initial phase angle.
Now, assume that this electron interacts with 
a  linearly polarized E\&M wave propagating in the positive $z$-direction:
$E_x(z,t) = E_1 \cos(kz - \omega t)$, 
$E_y=E_z=0$, $B_y(z,t) =  E_1(ck/\omega)\cos(kz - \omega t)$,
and $B_x = B_z = 0$. 
The first order linearized equation is
\begin{equation} 
m_e\frac{ d \mathbf{v}^{(1)}}{dt} =-e\left[ \mathbf{E} + \frac{\mathbf{v}^{(0)}}{c}
 \times  \mathbf{B} \right] - e \frac{\mathbf{v}^{(1)}}{c} \times  \mathbf{B}_0 \mathrm{.} \nonumber \label{eq:first}
 \end{equation}
Expanding the momentum equation of the second order in the $z$-direction 
and averaging it over $\phi_0$, we obtain  
\begin{eqnarray} 
 \frac{ d v_{z}}{dt} =   &-&\frac{k}{2}\frac{c k}{\omega} \left(\frac{e E_1}{m_e} \frac{v_p}{c} \right)^2
 \left( \frac{\sin(\alpha t)}{ \alpha^2} -\frac{\cos(\alpha t)}{ \alpha} t\right) \nonumber \\
&+& \left(\frac{e E_1}{m_e}\right)^2 \frac{1-\beta_{\parallel}}{c}\left( \frac{\sin(\alpha t)}{\alpha}\right) \mathrm{,} \label{eq:z3}
\end{eqnarray}
where $\alpha = k v_{0z} - \omega + \omega_{ce}$ and $\beta_{\parallel} = v^{(0)}_{z}/c$. Eq.~(\ref{eq:z3}) is exact to the second order in $E_1$. 
The first and the second term of the right hand side are 
from $(\mathbf{v}^{(0)}/c) \times  \mathbf{B} $ and the third term  is 
from $(\mathbf{v}^{(1)}/c) \times  \mathbf{B}_0$ in Eq.~(\ref{eq:first}).  
The resonance condition, $\alpha = 0$, leads to $ \omega-k v_{0z} =  \omega_{ce}$,
and the resonance electron velocity is $v_{r} = (\omega-\omega_{ce})/ k $.
Note that $v_r$ is positive (negative) if $\omega > \omega_{ce} $
($\omega < \omega_{ce} $). 

\begin{figure}
\scalebox{0.6}{
\includegraphics{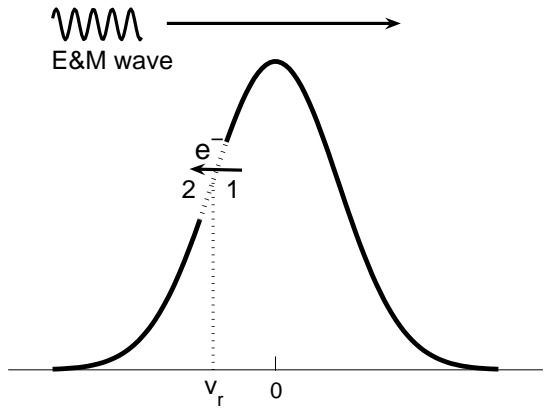}}
\caption{\label{fig1}
An electron emits a photon in the positive $z$-direction as it transitions from 1 to 2.  
See more details in the text.
}
\end{figure}

\begin{figure}
\scalebox{0.6}{
\includegraphics[angle=270]{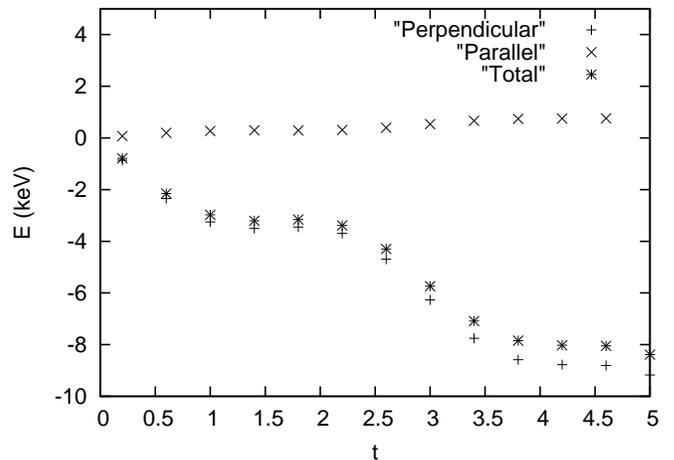}}
\caption{\label{fig3}
The energy of an electron in the presence of the magnetic field and the E\&M wave. The magnetic field is 1 T so that $\omega_{ce}= 0.025$ THz, and  $\omega = ck = 1/1.1 \  \omega_{ce} $. The intensity of the E\&M field is $ 2  \ \mathrm{MW} / \mathrm{cm}^2 \sec$. An electron has the initial perpendicular energy 80 keV and parallel energy  4.05 keV.
The x-axis is the time normalized by $(ck)^{-1}$ and the y-axis is the change in  the  parallel, perpendicular and total energy of the electron in the unit of keV. In this particular simulation, the electron parallel kinetic energy increases while its total (perpendicular)  kinetic energy decreases.  
}
\end{figure}

Eq.~(\ref{eq:z3}) shares similarity with the Landau damping analysis 
for a Langmuir wave~\citep{stix, songamma}.
From Eq.~(\ref{eq:z3}), we obtain 
 the kinetic energy loss rate of electrons over the electron distribution in the limit of $\alpha t > 1$~\citep{stix}:
\begin{equation} 
\frac{d \epsilon}{d t} =  \frac{\pi}{2}  \frac{E_x^2}{4\pi}\frac{\omega_{pe}^2}{k^2} 
\left[ \frac{ck}{\omega} \langle \frac{ \beta_{\perp}^2 }{2}  \frac{d f}{d v}\rangle  -   \langle \frac{f(v)}{c} (1-\beta_{\parallel})\rangle \right] \label{eq:Landau}
   c k  \mathrm{.}
\end{equation}
where $\beta_{\perp}^2 = v_p^2/c^2  = (v_x^2 + v_y^2)/c^2$, $\beta_{\parallel} = v_z / c$,   
$\omega_{pe}^2 = 4 \pi n_e e^2 / m_e $ is the plasma frequency, 
$f$ is the electron distribution function with the normalization of $\int f d^3v   = 1$, 
and $\langle \rangle$ is the ensemble average with $v_z = v_r$ or $\langle A \rangle =  \int   A\delta(v -v_r) d^3 v $. 
By employing the full relativistic equation~\citep{chu} and taking the appropriate classical limit, we also verify that 
Eq.~(\ref{eq:Landau}) is correct in the energy conservation 
if  the exchange of the electron parallel and perpendicular energy 
with the E\&M wave is fully taken into account.
It should be noted that this equation of the energy exchange is well-known in the maser theory~\citep{chu}. 
The case when $v_r<0$  also seems  contradictory 
as the E\&M wave  (the electron)  gains (gains) the momentum (the parallel kinetice energy). 
This can be resolved by considering the perpendicular electron motion.  
The electron kinetic energy in the perpendicular direction acts as an energy storage, by which the energy difference
between the electron kinetic energy in the $z$-direction and the energy gain of the E\&M mode can be accounted for. 
We demonstrate this idea by a single particle simulation (Fig.~\ref{fig3}). 
As the resonant interaction between the electron and the wave progresses, 
the electron loses the momentum in the $z$-direction and, at the same time, gains the kinetic energy in the same direction.  
However, simultaneously, the electron loses more energy in the perpendicular direction. 
The ratio of the perpendicular energy loss to the parallel energy gain is roughly  $c/v_r$,
which is consistent with the momentum and the energy relation in the quantum mechanics; While the electron gains the kinetic energy in the $z$-direction, it loses the energy in the perpendicular direction
by transitioning from a higher energy Landau level to a lower one. 

Assuming the energy density of the wave is given 
as $\epsilon_{w} = \zeta E_x^2/4\pi $, we 
arrive at the wave instability growth rate (Landau growth rate) by equation the kinetic energy loss rate to the wave growth rate, $\Gamma \epsilon_{w} = (d \epsilon/d t)$: 
\begin{equation} 
\Gamma =\frac{1}{\zeta} \frac{\pi}{2} \frac{\omega_{pe}^2}{k^2} 
\left[ \frac{ck}{\omega} \langle \frac{ \beta_{\perp}^2 }{2}  \frac{d f}{d v}\rangle  -   \langle \frac{f(v)}{c} (1-\beta_{\parallel})\rangle \right] \label{eq:landau}
   c k  \mathrm{.}
\end{equation}
In the case of a Langmuir wave,
no instabilities exist if the electron distribution 
is peaked at $v=0$ and monotonically decreases with $v$
because the wave always sees the negative slope of the electron distribution 
at the resonance. 
However, this is no longer the case for the E\&M mode.
If $ \omega  < \omega_{ce}$, $v_r$ and $k$ can be of the opposite signs
and an {\em amplification} of the E\&M wave (as well as the damping) can occur.
An E\&M mode propagating rightward interacts resonantly with the electrons
of certain negative velocity (Fig.~\ref{fig1}), and extract 
the momentum from  the resonant electrons. 
The second term on the right-hand side of Eq.~(\ref{eq:landau}) is always a damping term, 
but the first term could be an amplification or a damping depending on the sign of $df/dv$.

If $\zeta$ is positive, 
 Eq.~(\ref{eq:landau}) predicts that the E\&M wave becomes unstable 
in a Maxwellian plasma with $ T_e > m_e c^2/2$, seemingly contradicting the Gardner's constraint~\citep{gardner, gardner2} which 
states that a plasma of an isotropic and monotonically decreasing distribution is stable. 
This apparent contradiction can be resolved with a proper estimation 
of the free energy in the plasma. 
If an E\&M wave is present in a plasma, the plasma  
acquires some free energy from the wave oscillation, and   
the E\&M wave can extract this free energy up to
the maximum intensity imposed by the Gardner's constraint.
More specifically, let us assume that the electron distribution is Maxwellian for $t < 0$,
and that an external E\&M wave of $E_x$ is suddenly turned on at $t = 0$. 
For $t>0$, the electrons acquire an additional kinetic energy of
$ \delta \epsilon =  \zeta E_x^2 /4 \pi  - \epsilon_{w} =  (\zeta - 0.5 (1 + (\omega/ck)^2)) E_x^2/4\pi$,
where $\epsilon_{w} = (1/8\pi)E_x^2(1 + (\omega/ck)^2)$ is the E\&M field energy density.
Let us assume that the E\&M wave grows from $E_x$ to $E_{fx} $ due to the instability. Then, 
the free  energy drained into the E\&M field energy is given as
 $ \delta \epsilon_{w} = 0.5 (1 + (\omega/ck)^2) (E_{fx}^2- E_{x}^2)/4\pi$. 
We obtain the maximum wave intensity  from the 
condition $  \delta \epsilon_{w} < \delta \epsilon $,
\begin{equation}
E_{fx}^2 < \frac{2\zeta}{ 1 + \omega^2/(ck)^2} E_{x}^2  \mathrm{.} \label{eq::landau}  
\end{equation}


\section{Landau theory: Fully Relativistic electrons}
While Eq.~(\ref{eq:landau}) is mathematically correct,
the instability condition for the Maxwellian plasma predicted by the theory is 
 $ T_e > m_e c^2/2$, so that the  relativistic consideration is necessary. 
The momentum equation for a relativistic electron is
\begin{equation} 
m_e\frac{d \gamma \mathbf{v}}{dt}= -e\left[ \mathbf{E} + \frac{\mathbf{v}}{c} \times  \mathbf{B} \right] - e \frac{\mathbf{v}}{c} \times  \mathbf{B}_0 \mathrm{,} \label{eq:rmomentum}
\end{equation}
where $\gamma^{-2} =  1 - v^2/c^2$ is the relativistic factor.
Following the same steps as in the classical case for a linearly polarized E\&M wave but retaining only the resonance term,
the electron energy loss rate after the average over the initial phase is
\begin{eqnarray} 
  \frac{ d (\gamma )}{dt} &=&   - \frac{1}{2 \gamma c^2}\left(\frac{e E_1}{ m_e} \frac{v_p}{c} \right)^2 \nonumber \Omega   \left( \frac{\sin(\alpha t)}{ \alpha^2} -\frac{\cos(\alpha t)}{ \alpha} t\right) \nonumber  \\ \nonumber \\ 
   &+&\frac{1}{\gamma c^2} \left(\frac{e E_1}{m_e}\right)^2  
\frac{\sin(\alpha t)}{\alpha}
\left(1-\frac{\beta_{\perp}^2}{2} - \frac{ck}{\omega}\beta_{\parallel}\right) 
\mathrm{,}  \label{eq:z5} \\ \nonumber 
\end{eqnarray}
where $\Omega(\omega, k, \beta_{\parallel})=  
(c^2k^2/\omega)  (1- \omega \beta_{\parallel}/ck)  - \omega_{ce}/\gamma(\beta)$,
and $\alpha = k v_z - \omega + \omega_{ce} / \gamma$. 
Eq.~(\ref{eq:z5}) is exact to the second order in $E_1$. 
In the regime where the gyro-frequency is very high so that $\alpha t \gg 1$, the following approximation can be used~\citep{stix}: 

\begin{eqnarray} 
\int \left[\frac{\cos(\alpha t)}{ \alpha} g \right]d^3 \mathbf{x}&\cong& 0 \mathrm{,}
\nonumber \\  \nonumber \\ 
\int  \left[ \frac{\sin(\alpha t)}{ \alpha^2} g \right]d^3 \mathbf{x} &\cong&  
    \int    \left[\frac{dg}{d \alpha} \pi \delta(\alpha) \right] d^3 \mathbf{x}  \mathrm{,} 
\label{eq:a} \\ \nonumber  \\ \nonumber 
\int \left[ \frac{\sin(\alpha t)}{ \alpha} g \right]d^3 \mathbf{x}&\cong& 
\int \left[  g \pi \delta(\alpha)\right] d^3 \mathbf{x} \mathrm{,} 
\end{eqnarray}  
where $g$ is a smooth function of the velocity $\mathbf{\beta} = \mathbf{v} / c$, $dg/d\alpha = (\nabla_{\beta}\alpha \cdot\mathbf{ \nabla}_{\beta} g/|\mathbf{\nabla} \alpha(\beta)|^2)$ is the derivative of the $g$ in the direction of the gradient of $\alpha$.   
Employing the energy conservation, the growth rate of the E\&M wave can be obtained by equating the energy growth rate of the E\&M wave to the kinetic energy dissipation rate~\citep{stix}
\begin{equation} 
  \Gamma \frac{E_1^2}{ 8 \pi} \cong\langle   \frac{ d (\gamma )}{dt} m_e c^2 \rangle_{AVG} \mathrm{,} \label{eq:z10}
\end{equation} 
where $\langle \rangle_{AVG} $ is the ensemble average over the electron distribution. For  $\alpha t \gg 1 $, the growth rate $\Gamma$  is given from Eqs.~(\ref{eq:z5}), (\ref{eq:a}) and (\ref{eq:z10}) as~\citep{stix}
\begin{eqnarray} 
\Gamma  = &+& \frac{1}{\zeta} \left[\frac{\pi}{2} \frac{\omega_{pe}^2}{c^2k^2} 
\langle \frac{\mathbf{\nabla}_{\beta}S }{|\mathbf{\nabla} S(\beta)|^2 } \cdot\mathbf{ \nabla}_{\beta}\left( f  \frac{\Omega(\omega, k, \beta_{\parallel})}{\gamma} \frac{\beta_{\perp}^2}{2}  \right)   \rangle_{S=0} \right]  \nonumber \\
 &-& \left[\frac{\pi}{2} \frac{\omega_{pe}^2}{c^2k^2}   
\langle   \frac{1-\frac{\beta_{\perp}^2}{2} -\frac{ck}{\omega} \beta_{\parallel}}{\gamma} f \rangle_{S=0}\right] ck \mathrm{,} \label{eq:landau2} \\  \nonumber 
\end{eqnarray}
where $\mathbf{\beta} = \mathbf{v} / c$, and $S(\mathbf{\beta}) = \beta_{\parallel} - \omega / ck  + \omega_{ce} / (ck \gamma(\beta))$,
$f$ is the electron distribution with the normalization
of   $\int f d^3 \mathbf{\beta} = 1$, and 
$\langle A \rangle_{S=0} = \int \delta(S) A d^3 \mathbf{\beta} $ is the integration of the velocity space with the constraint $S=0$. 
Depending on whether more electrons lose or gain the energy in the resonance boundary $S(\mathbf{\beta}) = 0$, 
the growth rate, $\Gamma$ in Eq.(\ref{eq:landau2}),  could be of either sign.
Let us consider the classical limit of Eq.~(\ref{eq:landau2}),
$ S\cong \beta_{\parallel}- \omega/ck - \omega_{ce} / ck $ and $\gamma \cong 1$.  
The resonance boundary is the same as the case in Eq.~(\ref{eq:landau}),
and Eq.~(\ref{eq:landau2}) is simplified to 
\begin{eqnarray} 
\Gamma = &+&\frac{1}{\zeta} \frac{\pi}{2} \frac{\omega_{pe}^2}{k^2} 
\left[ \langle \frac{ \beta_{\perp}^2}{2}  \frac{d f}{d v}\rangle \right] \Omega 
\nonumber \\ 
&-&\frac{1}{\zeta} \frac{\pi}{2} \frac{\omega_{pe}^2}{k^2} \left[
   \langle \frac{f(v)}{c} (1-\frac{\beta_{\perp}^2}{2} -\frac{ck}{\omega}\beta_{\parallel})\rangle \right] ck  \label{eq:landau3}
   \mathrm{.}
\end{eqnarray}
where $\Omega = (c^2k^2/\omega)  (1- \omega \beta_{||}/ck)  - \omega_{ce}$.
Eqs.~(\ref{eq:landau2}) and (\ref{eq:landau3}) are the major result of our analysis.
While Eq.~(\ref{eq:landau2}) is similar with Eq.~(\ref{eq:landau}),
there is a crucial difference;
at $\omega = ck $, there could be an instability
from Eq.~(\ref{eq:landau}), which is not the case with Eq.~(\ref{eq:landau2}).
One of the key limitations of Eq.~(\ref{eq:landau}) is  the fact that
the change of the cyclotron resonance frequency 
due to the change in the relativistic electron mass
was not properly taken into account.

As the distribution function described by the Vlasov equation 
is incompressible in the canonical coordinate of $(p, q)$,
the Gardner's constraint is partially  relevant for the relativistic Maxwellian plasmas. 
Consider an initially isotropic Maxwellian plasma, which has no collective waves
at its initial state.
In the absence of the E\&M waves, the electron kinetic energy 
is a function of the canonical momentum, and,  due to the Gardner's constraint,
an certain amplified E\&M wave cannot escape from the plasma if the final plasma has no collective E\&M wave.
However, a steady-state plasma without any collective wave rarely exists.
Let us consider a plasma initially with some collective waves of appreciable energy 
(more specifically plasmons and photons with $\omega \cong \omega_{pe} $).
Noting that the Vlasov equation is compressible
in the coordinate of $(r, v) $ and the electron kinetic energy is given as 
$E = \sqrt{ m_e^2 c^4 + c^2(p - eA(r,t)/c)^2} $, 
 the Gardner's constraint no longer applies 
because the electron kinetic energy is now a function of the wave vector. 
Some collective E\&M waves 
could be radiated while some other waves still remain in the plasma. 
Eqs.~(\ref{eq:landau2}) and (\ref{eq:landau3}) suggest that 
the collective E\&M wave is a channel through which the plasma free energy
can be drained quickly. 

\section{ Instability}

One notable consequence of Eqs.~(\ref{eq:landau2}) and (\ref{eq:landau3})  is that
an unstable E\&M mode might  exist even in Maxwellian plasmas.
First, consider the case $ck < \omega $ in the 
classical limit given in Eq.~(\ref{eq:landau3}),
assuming the electron distribution is an isotropic Maxwellian 
with the electron thermal velocity $v_{th}$. 
In contrast to the example given in Fig.~\ref{fig1}, the first term 
on the right-hand side of Eq.~(\ref{eq:landau3}) is positive
if $v_r = \omega/k - \omega_{ce} / k  > 0$. For a positive $v_r$, the 
growth rate is given as
\begin{equation} 
\Gamma =\frac{1}{\zeta} \frac{\pi}{2}  \frac{\omega_{pe}^2}{c^2k^2}
\left( \frac{v_r}{ c^2}  |\Omega|  - (1-\beta_{th}^2 - \frac{ck}{\omega}\beta_r) k\right) f_{\parallel}(v_r) \mathrm{,}   
\end{equation}
where $\beta_{th} = v_{th} / c $, $\beta_r= v_r/c$, and 
$f_{\parallel}(v) = 1/\sqrt{2\pi} v_{th} \exp( - v^2/ (2 v_{th}^2)) $. 
The condition for the instability ($\Gamma > 0$)  is 
\begin{equation}
\label{eq:inst}
  \beta_r >  \frac{ 1-\beta_{th}^2}{ |\Omega|/ck  +ck/\omega} \mathrm{.}
\end{equation} 
For an instability to be appreciable,
$f(v_r)$ should be appreciably large or Eq.~(\ref{eq:inst}) should be satisfied
near $\beta_r \cong \beta_{th}$. 
The maximum possible growth rate at $v_r = v_{th} $ is given roughly as 
\begin{equation} 
\Gamma_{max} \cong  0.19 \times \frac{\omega_{\mathrm{pe}}^2}{ (ck)^2} \Omega \mathrm{,} \label{eq:max}
\end{equation}
where it is assumed that $\zeta \cong 1$. 
For an anisotropic plasma, the instability condition is
\begin{equation}
  \beta_r|\Omega| > \left( \frac{v_{\parallel} }{v_{\perp}}\right)^2
  \left[1- \left(\frac{v_{\perp}}{c}\right)^2 -\frac{ck}{\omega} \frac{v_r}{c}\right] \mathrm{,} 
\end{equation} 
where $v_{\parallel}$ ($v_{\perp}$) is the electron thermal velocity in the parallel (perpendicular) direction. 
If $v_{\parallel}/v_{\perp} \ll 1$, the instability range becomes much wider than the isotropic case,
and the maximum growth rate is given as 
\begin{equation} 
\Gamma_{max} \cong  0.19 \times \left(\frac{v_{\perp}}{v_{\parallel}}\right)^2 \frac{\omega_{\mathrm{pe}}^2}{ (ck)^2} \Omega \mathrm{.} \label{eq:max2}
\end{equation}

\begin{figure} 
\scalebox{0.3}{
\includegraphics{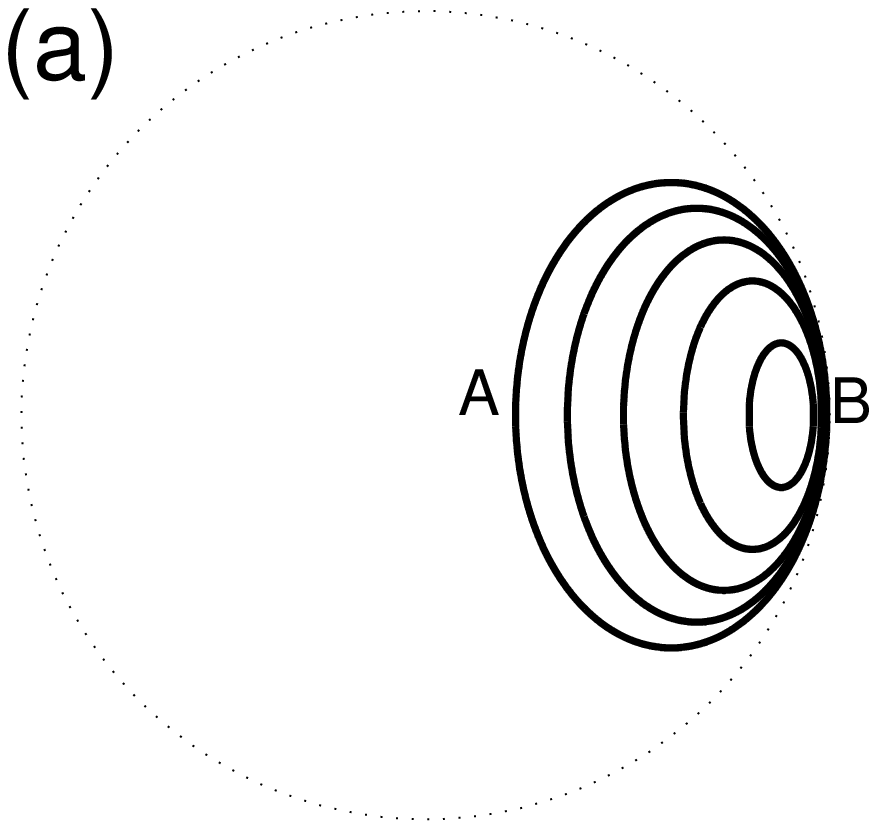}
\includegraphics{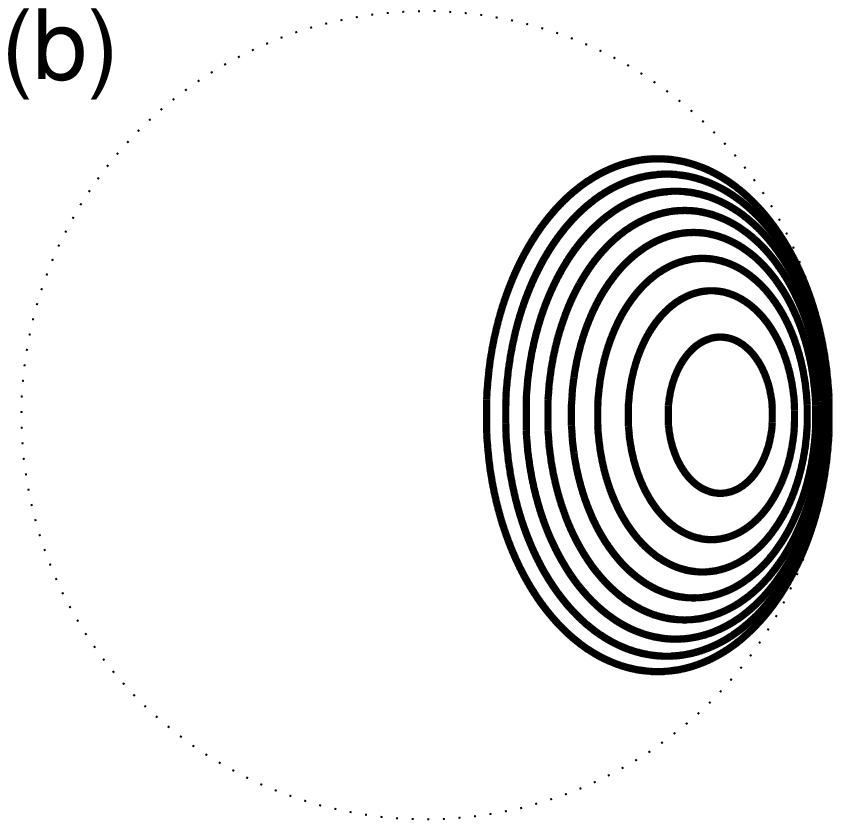}}
\caption{\label{fig2}
The contours of the resonance boundary ($S(\beta) = 0$) of the fully relativistic plasma
for the case of $\omega > ck > \omega_{ce}$.
(a) $\omega/(ck) = 1.1$ and $\omega_{ce}/(ck)$ decreases from 0.9 (the outer-most ellipse)
to 0.5 (the inner-most one) by 0.1, and
(b) $\omega_{ce}/(ck) = 0.9$ and $\omega/(ck)$ increases from 1.04 (the outer-most ellipse)
to 1.32 (the inner-most one) by 0.04.
Dotted lines are unit circles.
The abscissa is $\beta_{\perp}$ and the ordinate is $\beta_{\parallel}$.
}
\end{figure}

For the fully relativistic relation (Eq.~(\ref{eq:landau2})), 
the stability analysis gets more complicated.
Here we consider only one case of $\omega > ck >\omega_{ce}$.  
The Maxwellian electron distribution is given as $f(\beta) \cong \gamma^3 \exp(- \gamma/T_e) $, which is the so-called Maxwell-Juttner's distribution~\citep{jutt}. 
When $\gamma_0 = T_e/m_ec^2 > 1$, it is peaked at $\gamma = 3\gamma_0$ 
with the width of $\delta \beta \cong s/\gamma_0^2$. 
Here, we assume that, at the resonance, the derivative of the distribution $f$ is large dominating other terms. Then, Eq.~(\ref{eq:landau2}) can be simplified to 
\begin{eqnarray} 
\Gamma  = &+& \frac{1}{\zeta} \left[\frac{\pi}{2} \frac{\omega_{pe}^2}{c^2k^2} 
\langle   \frac{\Omega(\omega, k, \beta_{\parallel})}{\gamma} \frac{\beta_{\perp}^2}{2}  \frac{\mathbf{\nabla}_{\beta}S \cdot\mathbf{ \nabla}_{\beta} f }{|\mathbf{\nabla} S(\beta)|^2 }  \rangle_{S=0} \right]  \nonumber \\
 &-& \left[\frac{\pi}{2} \frac{\omega_{pe}^2}{c^2k^2}   
\langle   \frac{1-\frac{\beta_{\perp}^2}{2} -\frac{ck}{\omega} \beta_{\parallel}}{\gamma} f \rangle_{S=0}\right] ck \mathrm{.} \label{eq:landau4} \\  \nonumber 
\end{eqnarray}
Assuming $\beta \cong 1 $, the condition $\Gamma > 0 $ can be recasted as 
\begin{eqnarray}
\int_{S=0} \mathbf{d^3 \beta } \left(\frac{\mathbf{\nabla_{\beta}} S \cdot\mathbf{ \nabla_{\beta}} f}{|\nabla S(\mathbf{\beta})|^2 }\right)  \left(\frac{\omega_{ce} }{ \gamma ck } + 1\right) \left(1-\frac{ck}{\omega} \right) \nonumber \\ \nonumber   \\ 
>   \int_{S=0} \mathbf{d^3 \beta }   f(\beta_{res})\left(1-\frac{ck}{\omega}\beta_{res} \right) \mathrm{.} \label{eq:inst4} 
\end{eqnarray}
When  $\omega > ck >\omega_{ce}$,  
the resonance boundary is of an elliptical shape (Fig.~\ref{fig2}).
The resonance surface $S$ has the maximum
at $\beta_{max} = 1/ \sqrt{ 1 + (\omega_{ce}/ck)^2 } $. 
The necessary condition for the existence of the resonance is
$ \omega / ck < \beta_{max} + \omega_{ce}/\gamma_{max}$, where $\gamma_{max} = (1 -\beta_{max}^2)^{-1/2}$.
If the electron distribution has the relatively high slope near  the resonance region ($\gamma_{max} \cong  3 T_e / m_e c^2$), 
 Eq.~(\ref{eq:inst4}) is  possible since  $ |\nabla_{\beta} S | \ll 1$ near the resonance. 
The growth rate without the gyro-damping term can be estimated, using 
$\nabla_{\beta} f \cong \gamma^2(3-\lambda \gamma) f \mathbf{\beta}$   and $\beta \cong 1$, as  

\begin{equation}  \Gamma_{max} \cong  \frac{\omega_{\mathrm{pe}}^2}{ (ck)^2}
\int \delta(S) \left[\gamma \Omega_1 (3-\lambda) f \frac{ dS/d\beta   }{|\nabla_{\beta} S|^2}\right] d^3 \mathbf{\beta}   \mathrm{, } \label{eq:max8}
\end{equation} 
where  $(dS/d\beta) = (\mathbf{\beta} \cdot \nabla S)/|\mathbf{\beta}|$.   

A similar analysis with Eqs.~(\ref{eq:z3}) and (\ref{eq:z5}) is reported
in the context of the electron cyclotron maser instability~\citep{chu}; however, 
the major focus is in the short time growth rate ($\alpha t < 1$) when the magnetic field is not intense and/or the electron density is low. 
In the inertial confinement fusion or the astrophysical dense plasmas,  
the situation of the magnetic field of giga-gauss and the electron density
exceeding $10^{22} \mathrm{cm^{-3}} $ is common, and
the Landau damping analysis valid for $\alpha t \gg 1 $
should be sought instead of the conventional approach~\citep{chu,Treumann};
Eqs.~(\ref{eq:landau2}) and (\ref{eq:landau3}) provide a proper
estimation of the collective instabilities of  dense strongly magnetized plasma. 
While the instability analysis in this paper focused on Maxwellian plasmas, 
Eqs.~(\ref{eq:landau2}) and (\ref{eq:landau3}) apply to not only Maxwellian plasmas 
but also general relativistic strongly magnetized plasmas such as jets, shock regions and accretion disk where the non-Maxwellian electron distribution is common. 

An E\&M mode from the background noise could grow with the rate given in Eq.~(\ref{eq:landau}).
This radiation could be more explosive than the incoherent cyclotron radiation. 
The ratio of the cyclotron radiation power per electron $P$ to the electron kinetic energy is
$ \Gamma_{ci} = (2 P/ m v_p^2) =  (4/3) ( k e^2/m_e c^2) \omega$,
where the electrons are assumed to be non-relativistic.
Using Eq.~(\ref{eq:max}), we arrive at
$\Gamma_{max}/ \Gamma_{ci} \cong (n_e/k^3) (\Omega/\omega)$. 
If $ \Gamma_{max} / \Gamma_{ci} >1 $, the radiation from  unstable E\&M modes could be more explosive than the cyclotron radiation.  

\section{ Conclusion}

To summarize, the coherent E\&M  instabilities 
of strongly magnetized relativistic electrons are analyzed
in the context of the Landau damping theory. 
Our main results are Eqs.~(\ref{eq:landau2}) and (\ref{eq:landau3}).
Our theory based on the long-time scale is relevant with the hard x-ray and
the gamma ray generation in the strongly magnetized  dense plasmas.
In an astrophysical plasma,
a strong and large spatial scale magnetic field in the range of $10^{10} - 10^{15}$ gauss
is often encountered.
Its radiation may have to be re-examined in terms of the possible unstable modes identified in our analysis.
More specifically, the theory would be relevant with the generation of the soft and the hard x-ray,
and the gamma ray~\citep{gamma}.
The complications when applying the above theory to astrophysical plasmas are
the relativistic effect and the electron quantum diffraction~\citep{sonpla, sonprl, sonbackward, sonlandau}. 
One more relevant phenomenon is the non-inductive current drive~\citep{Fisch, sonprl}.   
During the E\&M wave amplification, 
the electrons giving the energy to the E\&M wave
lose their momentum at an increased rate, 
as the collision frequency increases
compared to the case of no interaction with the E\&M wave. 
This leads to the well-known non-inductive current drive~\citep{Fisch, sonprl}. 

The authors are thankful to Dr.~I.~Dodin and Prof.~N.~J.~Fisch for many useful
discussions and advice.

\bibliography{tera2}

\end{document}